\documentclass{ws-ijmpa}
\usepackage{graphicx}
\usepackage{amsmath}
\usepackage[latin1]{inputenc}
\usepackage{amsbsy}
\usepackage{color}
\usepackage{epsfig}
\usepackage{graphicx}
\usepackage{amsmath}
\usepackage{amssymb}
\usepackage{bbm}
\usepackage{amsbsy}
\usepackage{slashed}
\DeclareMathOperator{\tr}{tr}

\usepackage{xcolor}

\newcommand{\slsh}[1]{{\not \! #1}}

\usepackage[normalem]{ulem}  %\sout{}

\begin{document}
\title{Critical end point in a thermo-magnetic nonlocal NJL model}

\author{F. M\'{a}rquez}
\address{Santiago College, Avenida Camino Los Trapenses 4007, 
Lo Barnechea, Santiago, Chile. }
\author{R. Zamora}
\address{Instituto de Ciencias B\'asicas, Universidad Diego Portales, Casilla 298-V, Santiago, Chile.}
\address{Centro de Investigaci\'on y Desarrollo en Ciencias Aeroespaciales (CIDCA), Fuerza A\'erea de Chile, Santiago, Chile.}

\maketitle

\begin{abstract}
In this article we explore the critical end point in the $T-\mu$ phase diagram of a thermomagnetic nonlocal Nambu--Jona-Lasinio model in the weak field limit. We work with the Gaussian regulator, and find that a crossover takes place at $\mu, B=0$. The crossover turns to a first order phase transition as the chemical potential or the magnetic field increase. The critical end point of the phase diagram occurs at a higher temperature and lower chemical potential as the magnetic field increases. This result is in accordance to similar findings in other effective models. We also find there is a critical magnetic field, for which a first order phase transition takes place even at $\mu=0$.
\end{abstract}

%% \linenumbers

%% main text

\section{Introduction}

The main concern of this article is to study the critical end point (CEP) of the QCD phase diagram. Hadronic matter exists in either a chirally symmetric phase or a phase were chiral symmetry is spontaneously broken. The transition between these phases can be a first order or second order phase transition, or even a crossover. The type of transition occurring depends on the chemical potential and temperature at which it occurs. At low $\mu$ (and high $T$) the transition is either a crossover or a second order phase transition. At high $\mu$ (and low $T$) the transition is a first order one. The $(\mu, T)$ point separating both transitions in the phase diagram is called a CEP. The existence of the CEP in QCD was suggested a few decades ago \cite{Asakawa:1989bq,Barducci:1989wi,Barducci:1989eu,Barducci:1993bh}. To determine the position of the CEP in the phase diagramm one must work within the realm of nonperturbative QCD. Therefore, a number of different approaches are used to investigate this, namely lattice QCD \cite{Fodor:2004nz,Fodor:2001pe}, the linear sigma model \cite{renato2}, the Nambu--Jona-Lasinio (NJL)  model \cite{Costa:2008gr,Avancini:2012ee,Costa:2010zw,Costa:2007ie} and its nonlocal variant (nNJL) \cite{Scoccolannjl,Contrera:2010kz}.\\

More recently, the study of the QCD phase diagram in the presence of a magnetic field has been addressed in many articles \cite{Boomsma01,Loewe1,Agasian,Fraga,Fraga2,Andersen,Nosotros01,renato1,renato2,renato3,renato4,renato5,Gamayun,Marquez:2016fvb}. The magnetic field has been shown to have an effect on both the order of the phase transition and the critical temperature and chemical potential at which it occurs \cite{renato2}. Therefore, the magnetic field will have an effect on CEP position. Such a scenario may be found in heavy ion collisions, where a magnetic field is produced in presence of hadronic matter \cite{heavy01}.\\

The NJL model was originally proposed as model of interacting nucleons \cite{Nambu1,Nambu2} and later reinterpreted as a model of interacting quarks \cite{Volkov,Hatsuda}. The nNJL model was then introduced as way of including confinement in the model \cite{Birse01,Birse02,Fede02}. The nNJL has also shown good agreement with lattice data \cite{Scoccolannjl,Contrera:2010kz}. Therefore, the nNJL model is not only used to study confinement, but also to study nonpertubative properties of QCD. We will use this model in presence of a homogeneous magnetic field, in order to study the behavior of the CEP under such conditions.\\

The article is organized as follows. In Sec. 2, the model is presented and the uniform magnetic field is introduced. In Sec. 3 our results are presented and in Sec. 4 we discuss our conclusions and final remarks.\\

\section{Thermo-magnetic \lowercase{n}NJL Model.}

The nNJL model is described through the Euclidean Lagrangian
\begin{equation}\mathcal{L}_E=\left[\bar{\psi}(x)(-i\slashed{\partial}+m)\psi(x)-\frac{G}{2}j_a(x)j_a(x)\right],\label{lagrangiannjl}\end{equation}
with $\psi(x)$ being the quark field. The nonlocal aspects of the model are incorporated through the nonlocal currents $j_a(x)$
\begin{equation}j_a(x)=\int d^4y\,d^4z\,r(y-x)r(z-x)\bar{\psi}(x)\Gamma_a\psi(z),\end{equation} 
where $\Gamma_a=(1,i\gamma^5\vec{\tau})$ and $r(x)$ is the so-called regulator of the model in the configuration space. If $r(x)=\delta(x)$ then we would recover the original NJL model. It is usual to bosonize the model through the incorporation of a scalar ($\sigma$) and a pseudoscalar ($\vec{\pi}$) field. Then, in the mean field approximation,
\begin{eqnarray}
\sigma&=&\bar{\sigma}+\delta\sigma\\
\vec{\pi}&=&\delta\vec{\pi},
\end{eqnarray}
where $\bar{\sigma}$ is the vacuum expectation value of the scalar field, serving as an order parameter for the chiral phase transition. The vacuum expectation value of the pseudoscalar field is taken to be null because of isospin symmetry. Quark fields can then be integrated out of the model \cite{Scoccola02,Scoccola04} and the mean field effective action can be obtained.
\begin{equation}\Gamma^{MF}=V_4\left[\frac{\bar{\sigma}^2}{2G}-2N_c\int\frac{d^4q_E}{(2\pi)^4}\tr\ln S_E^{-1}(q_E)\right],\end{equation}
with $N_f=2$, the number of light-quark flavors and $N_c=3$, the number of colors in the model. Here, $S_E(q_E)$ is the Euclidean effective propagator
\begin{equation}S_E=\frac{-\slashed{q}_E+\Sigma(q_E^2)}{q_E^2+\Sigma^2(q_E^2)}.\label{Euprop}\end{equation}
Here, $\Sigma(q_E^2)$ is the constituent quark mass
\begin{equation}\Sigma(q_E^2)=m+\bar{\sigma}r^2(q_E^2).\end{equation}
Finite temperature ($T$) and chemical potential ($\mu$) effects can be incorporated through the imaginary time formalism (ITF) or Matsubara formalism. To do so, one can make the following substitutions
\begin{eqnarray}
V_4&\rightarrow&V/T\\
q_4&\rightarrow&-q_n\\
\int\frac{dq_4}{2\pi}&\rightarrow&T\sum_n,
\end{eqnarray}
where $q_n$ includes the Matsubara frequencies
\begin{equation}q_n\equiv(2n+1)\pi T+i\mu.\end{equation}
With this, the propagator in Eq. (\ref{Euprop}) will now look like
\begin{equation}S_E(q_n,\boldsymbol{q},T)=\frac{\gamma^4q_n-\boldsymbol{\gamma}\cdot\boldsymbol{q}+\Sigma(q_n,\boldsymbol{q})}{q_n^2+\boldsymbol{q}^2+\Sigma^2(q_n,\boldsymbol{q})}.\label{Euprop2}\end{equation}
It is worth noting that the propagator in Eq. (\ref{Euprop2}) has no singularities. Since there are no poles at some $p^2$, the definition of an effective mass for the particle with such propagator is not clear and therefore the quasiparticle interpretation cannot be made.\\

The $\sigma$ field will evolve with temperature. This evolution can be computed through the grand canonical thermodynamical potential in the mean field approximation $\Omega_{MF}(\bar{\sigma},T,\mu)=(T/V)\Gamma_{MF}(\bar{\sigma},T,\mu)$ \cite{Kapusta01}. Then the value of $\bar{\sigma}$ must be at the minimum of the potential where 
$\partial\Omega_{MF}/\partial\bar{\sigma}=0$, which means
\begin{equation}
\left.\frac{\bar{\sigma}}{G}=2N_cT\sum_n\int\frac{d^3q}{(2\pi)^3}
r^2(q_E^2)\tr 
S_E(q_E)
\right|_{q_4=-q_n}.
\label{gap}
\end{equation}
From this equation one can get the temperature evolution of $\bar{\sigma}$. All of the computations have been made in ITF. Similar derivations are readily available in the literature (see for example \cite{Scoccola02,Scoccola06}).\\

We are interested in studying the model coupled to a homogeneous magnetic field. The derivative in the Lagrangian (\ref{lagrangiannjl}) is replaced by a covariant derivative
\begin{equation}
D_\mu=\partial_\mu + ie_fA_\mu,
\end{equation}
where $A^{\mu}$ is the vector potential corresponding to a homogeneous external magnetic field $\boldsymbol{B}=|\boldsymbol{B}|\hat{z}$ and $e_f$ is the electric charge of the quark fields (i.e. $e_u = 2e/3$ and $e_d = -e/3$). In the symmetric gauge,
\begin{equation}
A^{\mu}= \frac{B}{2}(0,-y,x,0).
\end{equation}
The Schwinger proper time representation for the propagator in the Euclidean space is given by \cite{Schwinger}

\begin{multline}
S_E(q_E)=\int_0^\infty ds \frac{e^{-s(q_{\|}^2+q_\perp^2
   \frac{\tanh (eBs)}{eBs} + M^2)}}{\cosh (eBs)}
   \\\times\biggl[\left(\cosh (eBs) -i \gamma_1 \gamma_2 \sinh (eBs)\right)
   (M-\slsh{q_{\|}}) - \frac{\slsh{q_\bot}}{\cosh(eBs)} \biggr],
\end{multline}
with $q_\|^2=q_0^2+q_3^2$, $q_\bot^2=q_1^2+q_2^2$ and where $e$ is the charge of the particle and $B$ is the magnetic field.\\ 

For simplicity, we will consider the weak magnetic field case. By weak we mean that the magnetic field is weak with respect with the dominant energy scale in the problem, i.e. $eB<T^2$ or $eB < \mu^2$ \cite{mexicanos}. The Euclidean fermionic propagator in this region can be written as \cite{Taiwan}
\begin{eqnarray}
&&S_E(q_E) = \frac{(\Sigma(q_E^2)-\slsh{q_E})}{q_E^2+\Sigma^2(q_E^2)} - i\frac{\gamma_1 \gamma_2(eB)( \Sigma(q_E^2)-\slsh{q}_{E \|})}{(q_E^2+\Sigma^2(q_E^2))^2} \nonumber \\
&+&\frac{2 (eB)^2 q_{E \bot}^2}{(q_E^2+\Sigma^2(q_E^2))^4} \nonumber \\ 
&\times&\biggl[ (\Sigma(q_E^2)-\slsh{q}_{E \|}) 
+ \frac{\slsh{q}_{E \bot}(\Sigma^2(q_E^2)+{q}_{E \|}^2)}{q_{E \bot}^2} \biggr]. \label{debil}
\end{eqnarray}
\section{Results}

Throughout this work, we will use the Gaussian regulator for the nNJL model in the Euclidean momentum space, i.e.
\begin{equation}r^2(q_E)={\rm e}^{-q_E^2/\Lambda^2}.\end{equation}
For the parameters of the model, we take \cite{LoeweMorales} $m=10.5$ MeV, $\Lambda=627$ MeV and $G=5\times10^{-5}$ MeV$^{-2}$. With this set of parameters we have $\bar{\sigma}_0=339$ MeV.\\

The gap equation is solved for different values of the magnetic field and the chemical potential, obtaining the behavior of $\bar{\sigma}(T)$ for each pair of $(eB,\mu)$ values. This solution is found by numerical computation. This allows to determine the critical temperature for chiral phase transition, as well as the nature of the phase transition, namely if it is a first or second order phase transition, or rather a crossover.\\

\begin{figure}[!htb]
\begin{center}
\includegraphics[scale=0.6]{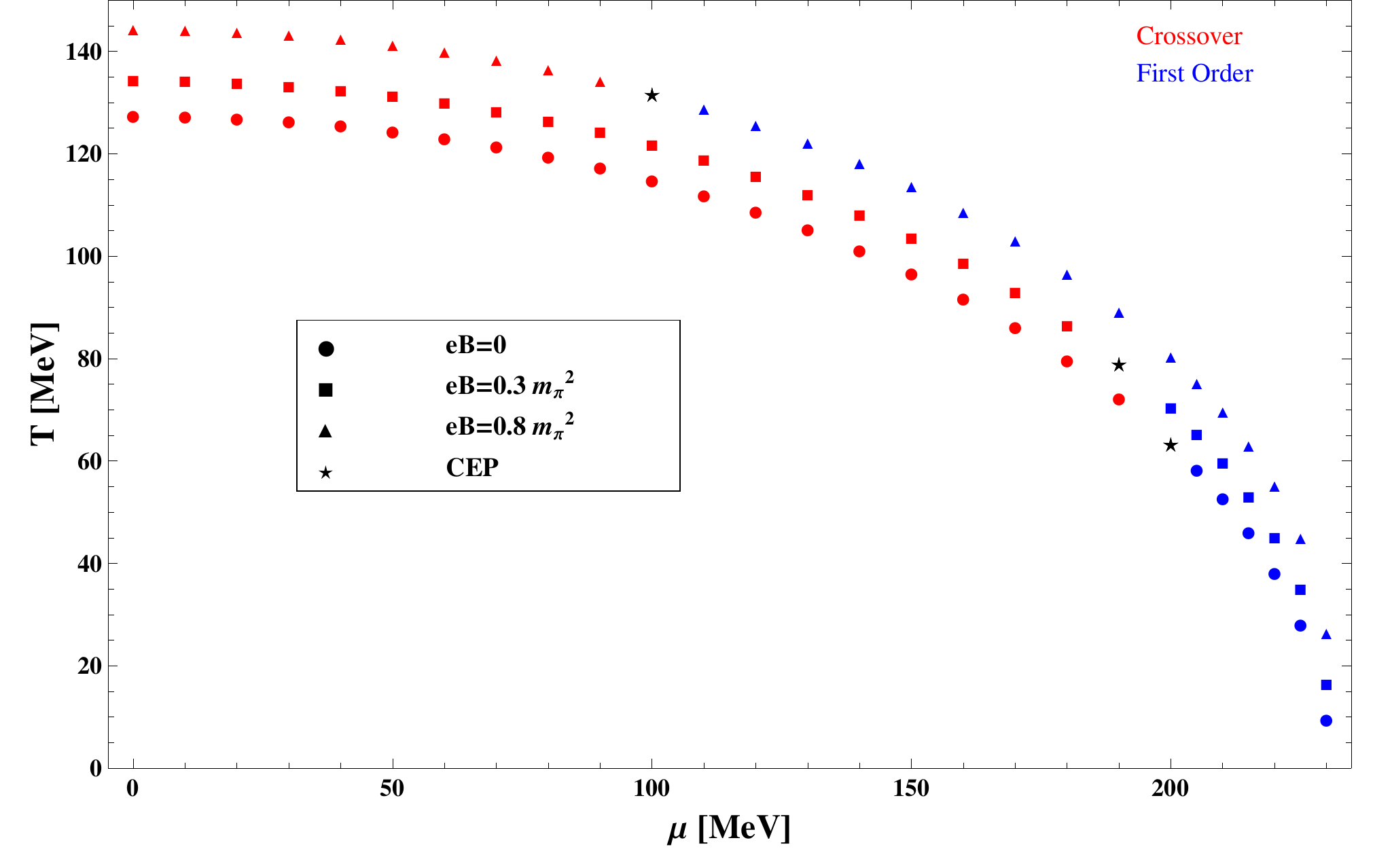}
\caption{$T-\mu$ phase diagram of the model, for different values of the magnetic field. Values of $eB$ are indicated in the figure. Red points signal a crossover while blue ones signal a first order phase transition. The stars show the position of the critical end point.}
\label{F1}
\end{center}
\end{figure}

Figure \ref{F1} shows the $T-\mu$ phase diagram for different values of the magnetic field in terms of the pion mass $m_\pi=134.97$ MeV \cite{valor}. In all cases, we initially have a crossover that, at high enough chemical potential, turns to a first order phase transition. We can also see that the CEP moves to the left of the phase diagram as the magnetic field increases. It is worth noting that an opposite behavior is found in \cite{inagaki}, indicating that the CEP behavior is a model-dependent phenomenon. However, similar results have been found in \cite{renato2}. This means that for higher magnetic fields, the critical temperature increases while the critical chemical potential decreases.\\

\begin{figure}[!htb]
\begin{center}
\includegraphics[scale=0.45]{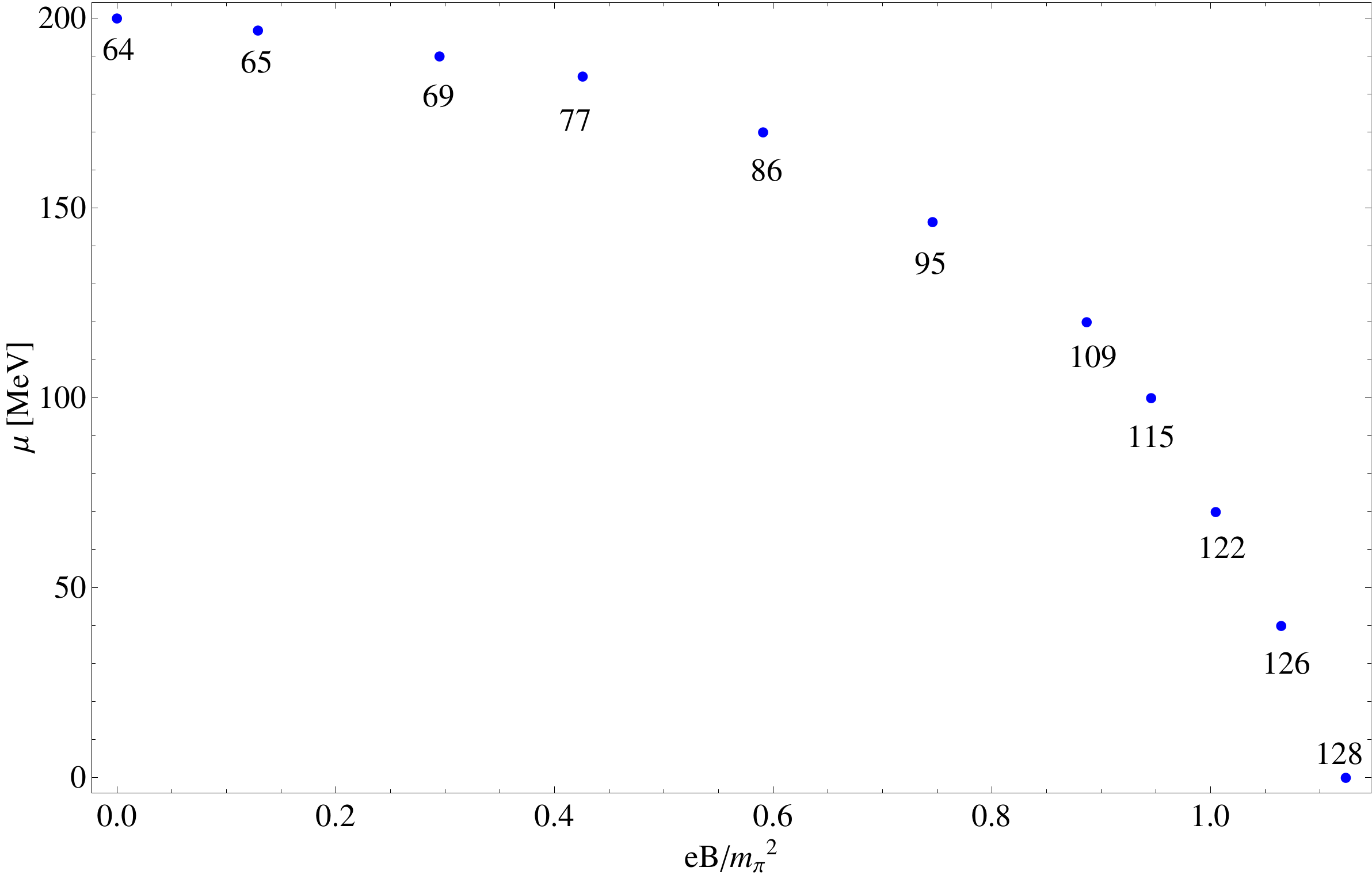}
\caption{Behavior of the chemical potential of the critical end point as a function of the magnetic field. The number below data points indicate the value of the temperature of the CEP for each case.}
\label{F3}
\end{center}
\end{figure}

Figure \ref{F3} shows the behavior of the chemical potential for the CEP as a function of the magnetic field. As can be seen from the figure, the chemical potential decreases as the magnetic field increases. Furthermore, at $eB=1.1 m_\pi^2$ MeV$^2$, the chemical potential for the critical endpoint is null, meaning that there is no longer a crossover in the phase diagram, but rather a first order phase transition at every critical temperature, therefore we can no longer define a CEP.\\

\begin{figure}[!htb]
\begin{center}
\includegraphics[scale=0.4]{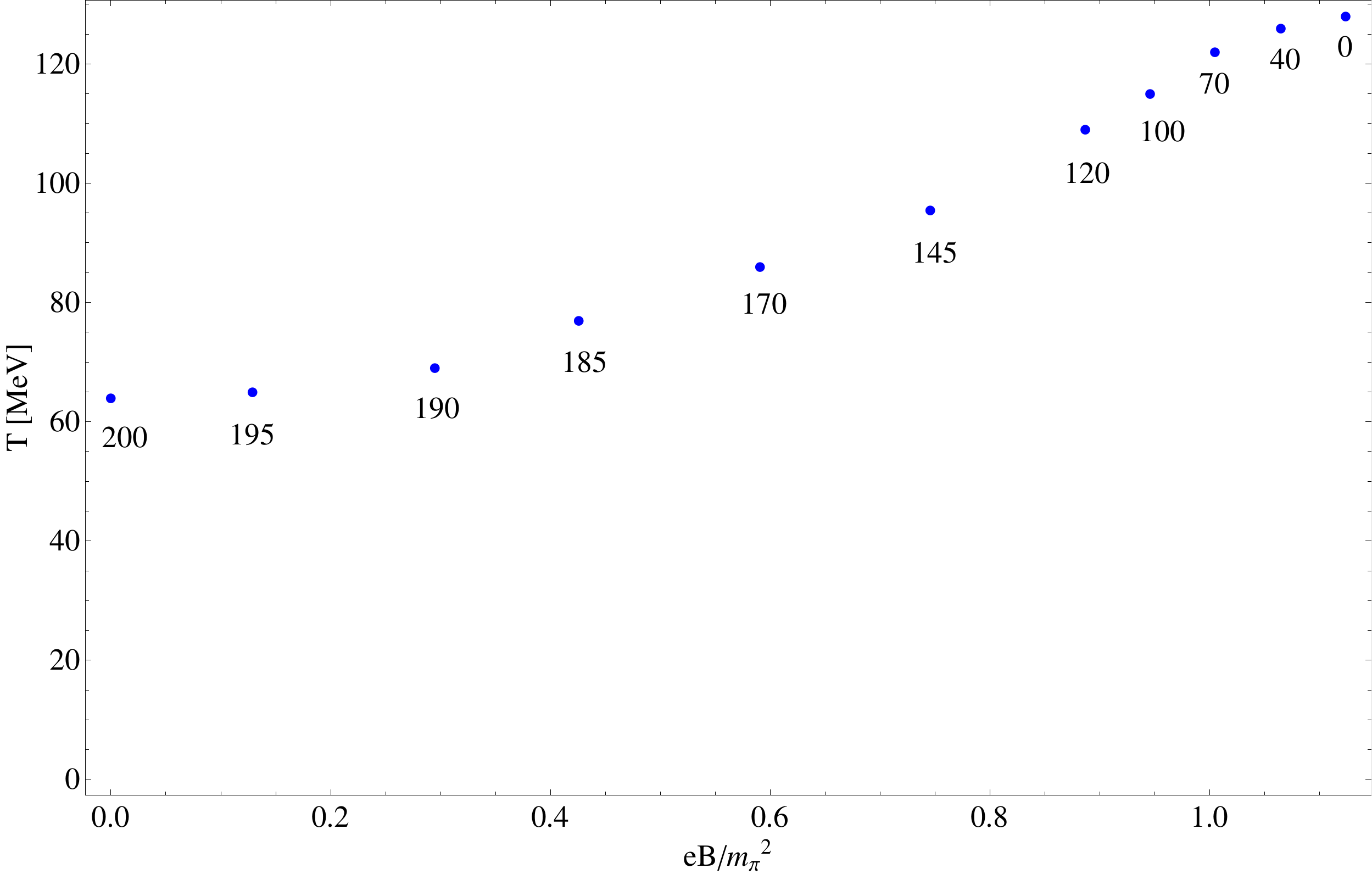}
\caption{Behavior of the temperature of the critical end point as a function of the magnetic field. The number below data points indicate the value of the chemical potential of the CEP for each case.}
\label{F2}
\end{center}
\end{figure}

Figure \ref{F2} shows the behavior of the temperature of the CEP as a function of the magnetic field. As can be seen from the figure, the temperature of the CEP increases as a function of the magnetic field. However, at $T=128$ MeV, the CEP can no longer be defined, as there is no crossover in the model.\\

\begin{figure}[!htb]
\begin{center}
\includegraphics[scale=0.4]{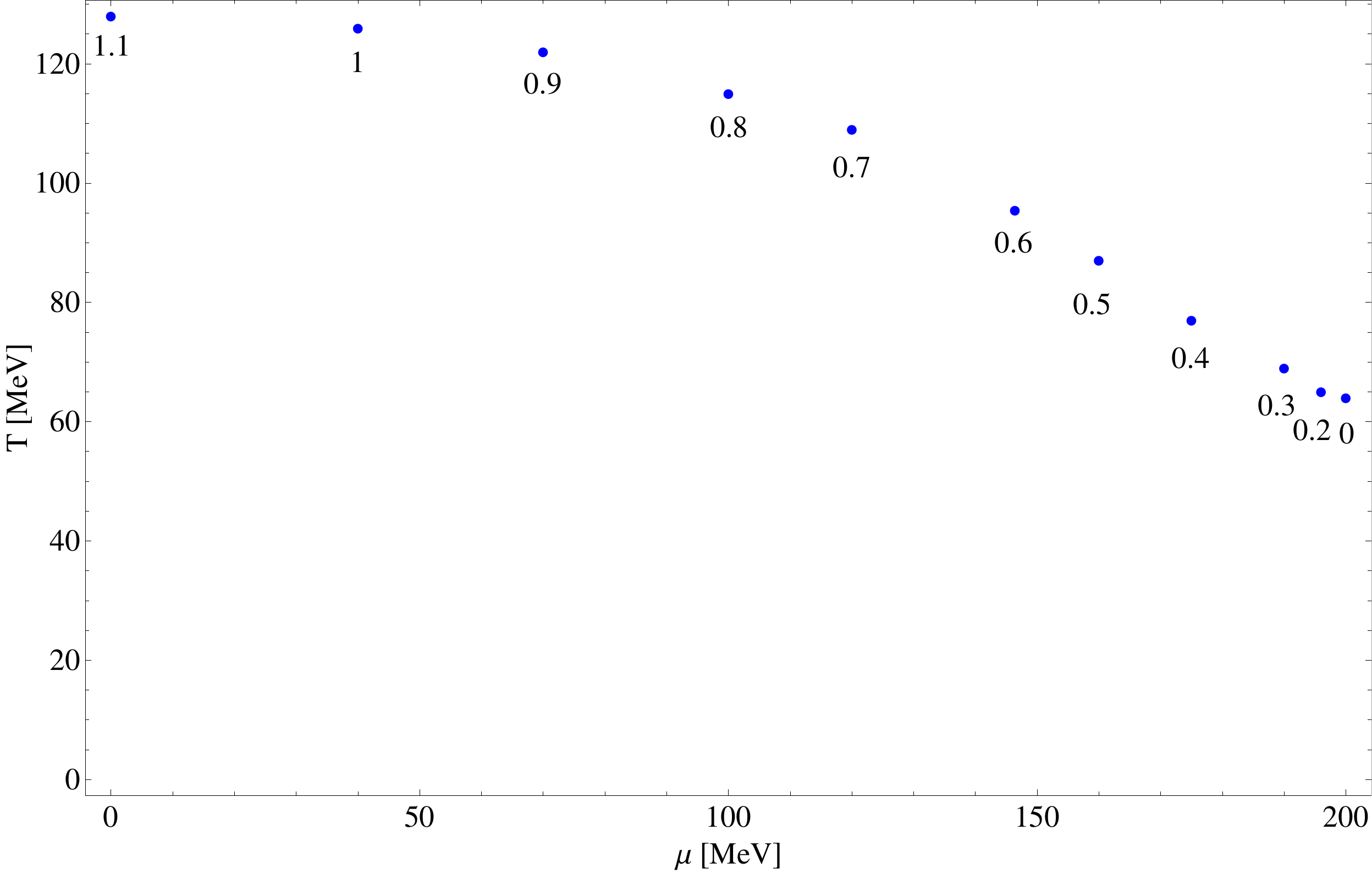}
\caption{Behavior of the temperature of the CEP as a function of the chemical potential of the CEP. The number below data points indicate the value of the magnetic field for each case. The values of the magnetic field are normalized by the pion mass.}
\label{F4}
\end{center}
\end{figure}

Figure \ref{F4} shows the behavior of the temperature of the CEP as a function of the chemical potential of the CEP. As the chemical potential increases, the temperature decreases.

\section{Conclusions}

In this article, we obtained the $T-\mu$ phase diagram for a thermo-magnetic nNJL model. We find that there is a chiral phase transition that can either be a crossover or a first order phase transition at $B=0$, depending on the value of the chemical potential. One can define a CEP as the set of $(T,\mu, eB)$ values that separate the crossover from the first order phase transition. We find that as the magnetic field increases the temperature of the CEP also increases, while the chemical potential of the CEP decreases. This is in agreement with results obtained in different effective models when working in the mean-field approximation. Furthermore, it has been shown \cite{renato4} that inverse magnetic catalysis will be found when going beyond mean field. We studied the beahvior of the CEP in the phase diagram. If the magnetic field is high enough ($eB>1.1 m_{\pi}^2$), the CEP will vanish, meaning that we will no longer have a crossover in our model, but rather a phase transition for any temperature. At this point is no longer possible to define a CEP. It is worth noting that, while this will occur at higher $B$, the magnetic field is still weaker than the dominant energy scale. Therefore, the weak field approximation is still valid in this region.

\section{Acknowledgements}

R. Zamora would like to thank support from CONICYT FONDECYT Iniciaci\'{o}n under grant No. 11160234.

%\section*{References}

\bibliography{BNJL}{}
\bibliographystyle{ws-ijmpa}

\end{document}